\documentclass[prl,twocolumn,showpacs,preprintnumbers]{revtex4}
\usepackage{graphicx}

\begin{document}

\title{First-principles calculations and bias-dependent STM measurements at the $\alpha $-Sn/Ge(111) surface: a clear indication for the 1U2D configuration}
\author{Paola Gori}
\altaffiliation[Also at ]{NAST Centre, Via della Ricerca Scientifica 1, 00133 Roma, Italy}
\email{gori@ism.cnr.it}
\author{Fabio Ronci}
\author{Stefano Colonna}
\author{Antonio Cricenti}
\altaffiliation[Also at ]{Department of Physics and Astronomy, Vanderbilt University, Nashville, Tennessee 37235-1807, USA.}
\altaffiliation[Also at ]{NAST Centre, Via della Ricerca Scientifica 1, 00133 Roma, Italy}
\affiliation{ISM CNR, Via del Fosso del Cavaliere 100,
I-00133 Roma, Italy.}
\author{Olivia Pulci}
\altaffiliation[Also at ]{NAST Centre, Via della Ricerca Scientifica 1, 00133 Roma, Italy}
\affiliation{ETSF, INFM-CNR Dipartimento di Fisica, Universit\`{a} di Roma ``Tor Vergata", Via della Ricerca Scientifica 1, I-00133 Roma, Italy.}
\author{Guy Le Lay}
\altaffiliation[Also at ]{UFR Sciences de la Mati\`{e}re, Universit\'{e} de Provence, Marseille, France.}
\affiliation{CINaM-CNRS, Campus de Luminy, Case 913, F-13288 Marseille Cedex 09, France}

\date{\today}

\begin{abstract}
The nature of the $\alpha $-Sn/Ge(111) surface is still a matter of debate. In particular, two possible configurations have been proposed for the 3$\times$3 ground state of this surface: one with two Sn adatoms in a lower position with respect to the third one (1U2D) and the other with opposite configuration (2U1D). 
By means of first-principles quasiparticle calculations we could simulate STM images as a function of bias voltage and compare them with STM experimental results at 78K, obtaining an unambiguous indication that the stable configuration for the $\alpha $-Sn/Ge(111) surface is the 1U2D. The possible inequivalence of the two down Sn adatoms is also discussed.
\end{abstract}

%\pacs{68.35.Ja, 68.35.Rh, 68.37.Ef}

\maketitle

The Sn/Ge(111) system has been thoroughly studied for its very rich phase diagram as a function of Sn thickness and sample thermal treatment \cite{Ichikawa,Gothelid}. In particular, the $\alpha $-phase is obtained by evaporating 1/3 ML of Sn on a clean c(2$\times $8) Ge(111) surface after a short annealing at about 500K. This surface is characterized by Sn adatoms regularly located on one out of three T$_4$ sites of the bulk terminated Ge(111) surface, resulting in a ($\sqrt{3}$$\times$$\sqrt{3}$)$R30{^\circ}$ reconstruction. A particular interest for this phase arose after the discovery of a gradual and reversible phase transition to a 3$\times$3 reconstruction below 220K \cite{Carpinelli}. This well known, but not yet completely understood, surface transition was initially explained as a Charge Density Wave (CDW) formation below a critical temperature\cite{Carpinelli,Zhang,Pedersen,Baddorf}. However, due to the absence of nesting at the Fermi surface \cite{Carpinelli,Chiang} and to the presence of two Sn-4d core level components with a 2:1 relative intensity ratio both above and below the transition temperature \cite{Uhrberg1,Uhrberg2}, this model has been questioned. Alternative models describing the $\sqrt{3}$$\times$$\sqrt{3}$ to 3$\times$3 transition as an order-disorder transition have been put forward. Among them, the ``dynamical fluctuation'' model \cite{Avila,Ronci}, suggests that the 3$\times$3 reconstruction is the ground state and the $\sqrt{3}$$\times$$\sqrt{3}$  reconstruction, observed at room temperature, results from thermally activated rapid vertical oscillations of the Sn atoms. Below the critical temperature the adatoms fluctuation is frozen in the 3$\times$3 periodicity in which Sn adatoms have different heights with respect to the underlying Ge substrate giving rise to a buckled surface.
\\
The exact structure of the 3$\times$3 reconstruction is still a matter of debate. In fact, two possible configurations of this surface have been proposed, one with two Sn adatoms in a higher position with respect to the third one (two adatoms in Up position and one in Down position, 2U1D for brevity hereafter) and the opposite configuration (1U2D). STM results alone cannot tell whether the surface configuration is 1U2D or 2U1D, because imaging empty or filled electronic states results in a honeycomb (an apparent 2U1D) image or a complementary hexagonal (an apparent 1U2D) one, respectively \cite{Carpinelli,Jurczyszyn}. In order to understand which configuration describes such a system, surface-sensitive structural techniques \cite{Okasinski,Bunk,Davila,Fukaya,Lee}, Sn-4d core level photoemission spectroscopy \cite{Uhrberg1,Uhrberg2,Tejeda}, non-contact AFM investigations \cite{Yi}, theoretical calculations\cite{Avila,Perez,Jurczyszyn,Ballabio,Melechko,Profeta} were used, producing conflicting results. \\
In this paper we present first-principles quasiparticle calculations that, combined with a bias-dependent STM study, clearly demonstrate a buckled surface structure for the $\alpha $-Sn/Ge(111) system at 78K with one out of three Sn adatom displaced upwards, solving the issue of the 1U2D versus 2U1D configuration. The inequivalence of the two down Sn adatoms is also discussed.
%
%\begin{figure}[htb!]
\begin{figure}[b]
\includegraphics[width=0.4\textwidth]{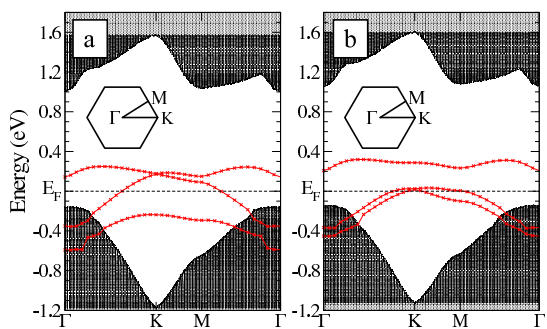}
\caption{\label{Fig1}(color online). GW-corrected surface band structure for the 1U2D (a) and 2U1D (b) configurations along high-symmetry directions of the 3$\times$3 SBZ (shown in the inset).}
\end{figure}
\\
First-principles calculations have been carried out at a first stage at the DFT level (in a plane-wave pseudopotential implementation
\cite{pwscf}) and in the Local Density Approximation (LDA). The Sn/Ge(111) surface has been represented in a repeated slab 
geometry consisting of six Ge layers of 9 atoms each, saturated by H atoms on the bottom layer and with Sn adatoms on top \cite{det_dft}. 
Quasiparticle corrections to DFT-LDA eigenvalues have been afterwards calculated in the GW approximation \cite{GW} to allow a closer 
comparison of theoretical and experimental results. It has been shown \cite{Gori} that LDA eigenvalues give already a good 
description of surface states introduced in the band gap by tin for Sn/Ge(111). However, in order to obtain an accurate picture 
of the whole metal/semiconductor interface, it is necessary to go beyond DFT since electron screening is largely space dependent 
in such an inhomogeneous system and implies sizeable quasiparticle effects to be taken into account for those energy structures 
which are associated to germanium \cite{det_gw}.
\begin{figure}[b]
\includegraphics[width=0.4\textwidth]{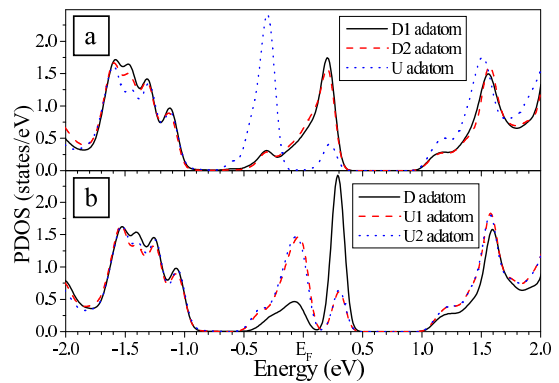}
\caption{\label{Fig2}(color online). Projected density of states at Up and Down tin adatoms for the 1U2D (a) and 2U1D (b) configurations.}
\end{figure}
STM images have been simulated using the Tersoff-Hamann model \cite{Tersoff}, as energy-integrated GW-corrected local density of states at a fixed height above the sample, using an average tip-sample distance of 5 \AA.
\begin{figure*}[htb!]
\includegraphics[width=0.9\textwidth]{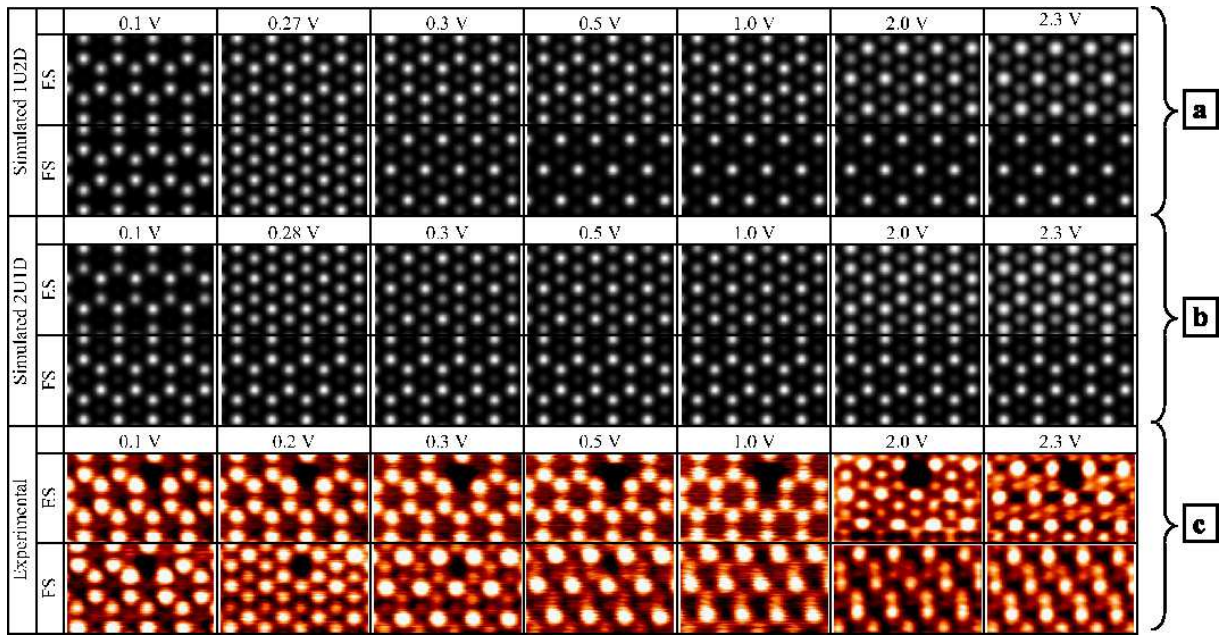}
\caption{\label{Fig3}(color online). Simulated 5$\times$3 nm STM images for the 1U2D (panel a) and 2U1D (panel b) configurations and experimental 5$\times$3 nm STM images (panel c, T = 78K, I = 20 pA) as a function of bias voltage. All the experimental images were obtained on the same area showing a Ge substitutional defect. ES= Empty States; FS = Filled States.}
\end{figure*}
Calculations were performed on the two possible 3$\times $3 reconstructions starting from initial configurations having a vertical 
buckling of 0.4 \AA\ for both 1U2D and 2U1D systems. In both cases, as expected \cite{Avila,Perez,Jurczyszyn}, the relaxation 
converged to the same energy minimum, corresponding to a 1U2D model with a vertical buckling of 0.36 \AA\ between up and down tin adatoms
\cite{num_lay}. Hence, we also considered a metastable 2U1D model having a vertical buckling of 0.20 \AA\ (a value close to the structural results reported in Ref. \cite{Lee}) for comparing the resulting band structure and STM images with the ones derived from the stable 1U2D configuration.
The GW-corrected surface band structures calculated along high-symmetry directions of the 3$\times $3 SBZ are reported in Figure \ref{Fig1} for the 1U2D and the 2U1D configurations. The three surface bands are associated to the three Sn dangling bonds: the ``up'' adatoms are characterized by a filled dangling bond while the ``down'' adatoms have partially occupied ones \cite{Degironcoli,Ortega}.
This is seen, for example, from the projected density of states (PDOS) at the three types of Sn adatoms reported in Figure \ref{Fig2}. Consequently, the two upper bands in Figure \ref{Fig1}a are associated to 2D adatoms and the third band at lower energy to 1U adatoms. Conversely, the 2U1D band structure reported in Figure \ref{Fig1}b shows that the two lower bands describe 2U adatoms and the third one located at higher energy describes D adatoms.
Such PDOS curves suggest that STM images obtained with very small bias voltages should mainly show the two D adatoms in the 1U2D case and the two U in the 2U1D case, resulting in both cases in a honeycomb pattern.
This hypothesis is confirmed by simulating STM images for the 1U2D and the 2U1D reconstructions (reported in Figure \ref{Fig3}a and \ref{Fig3}b, respectively) as a function of the bias voltage V. Indeed, in both cases the simulated images obtained at bias voltages lower than 0.2 V show a honeycomb pattern in both empty and filled states. Increasing the bias voltage, in the 1U2D case (Figure \ref{Fig3}a), the filled states images gradually revert to the expected hexagonal pattern, passing through an apparent $\sqrt{3}$$\times$$\sqrt{3}$ reconstruction at about 0.27 V, while the empty states images preserve the honeycomb pattern. As a result, the calculated images between 0.3 V and 1.0 V show the well-known complementary honeycomb and hexagonal patterns (for empty and filled states, respectively) reported in many papers. Interestingly enough, a further unexpected transition from honeycomb to hexagonal is observed in the empty states images at higher bias voltage. As a result, both the filled and empty states simulated images at 2.0 and 2.3 V show a hexagonal pattern.
The 2U1D simulated STM images reported in Figure \ref{Fig3}b show an opposite behavior: increasing the bias voltage above 0.2 V, the honeycomb to hexagonal transition occurs in the empty states images, crossing the apparently flat reconstruction at about 0.28 V, while the honeycomb pattern is maintained in the filled states images. Further increasing the bias voltage, a new hexagonal to honeycomb transition is observed in the empty states series, resulting in a honeycomb pattern for both empty and filled states images.
Summarizing the results obtained by analyzing the simulated STM images, we found that:
a) at low bias voltage (i.e. lower than about 0.2 V) for both the 1U2D and the 2U1D models STM images with honeycomb pattern are predicted in both empty and filled states; b) at intermediate bias voltage the simulated STM images of the 1U2D system confirm previous results obtained at $\pm\ 0.55 V$ \cite{Jurczyszyn} showing honeycomb and hexagonal patterns for empty and filled states images, while the 2U1D results give the opposite condition; c) at high bias voltage, calculations predict that the STM images, for both empty and filled states, should provide a picture of the true surface reconstruction (i.e. hexagonal for the 1U2D and honeycomb for the 2U1D).
\\
STM measurements were carried out at 78 K using a Low Temperature UHV-STM (Omicron LT-STM, base pressure 5$\cdot 10^{-11}$ mbar) and a tungsten tip cleaned in vacuum by electron bombardment. Germanium substrates were cut from Ge(111) n-type wafers. A clean c(2$\times $8) Ge(111) surface was obtained using a standard sputter-annealing procedure \cite{Ronci}. A nominal 1/3 ML Sn deposition was performed at RT, followed by sample annealing at about 200${^\circ}$C. The formation of the $\alpha $-phase $\sqrt{3}$$\times$$\sqrt{3}$ Sn/Ge(111) was checked by LEED and STM.
\\
It is well known since 10 years by now \cite{Carpinelli} that in STM measurements empty (filled) states images display a honeycomb (hexagonal) pattern. However, such an undisputed evidence has always been obtained, to our knowledge, using intermediate bias voltage (usually $\pm\ 1 V$).
Here, in order to verify the theoretical predictions and to discriminate between 1U2D and 2U1D configurations, we acquired a series of STM images on the same sample area ranging from $\pm 0.1 V$ to $\pm\ 2.3 V$, reported in Figure \ref{Fig3}c.
The evolution of such experimental STM images is strikingly similar to the one reported in the simulated STM images series for the 1U2D configuration (Figure \ref{Fig3}a). The first predicted transition from honeycomb to hexagonal at low bias voltage in the filled states series is clearly visible in the first three images collected in the 0.1 $\div $ 0.3 V range. In particular, the 0.2 V image shows an apparent $\sqrt{3}$$\times$$\sqrt{3}$ reconstruction, as predicted by the simulated STM image at 0.27 V. Furthermore, the second expected transition from honeycomb to hexagonal at high bias voltage in the empty states series was verified as well in the experimental empty states STM images.
\\
All the evidences described so far clearly indicate a remarkable match between simulations obtained from ab-initio calculations and experiments, pointing out a 1U2D configuration for the $\alpha $-Sn/Ge(111) surface.
%
%\begin{figure}[htb!]
\begin{figure}[b]
\includegraphics[width=0.4\textwidth]{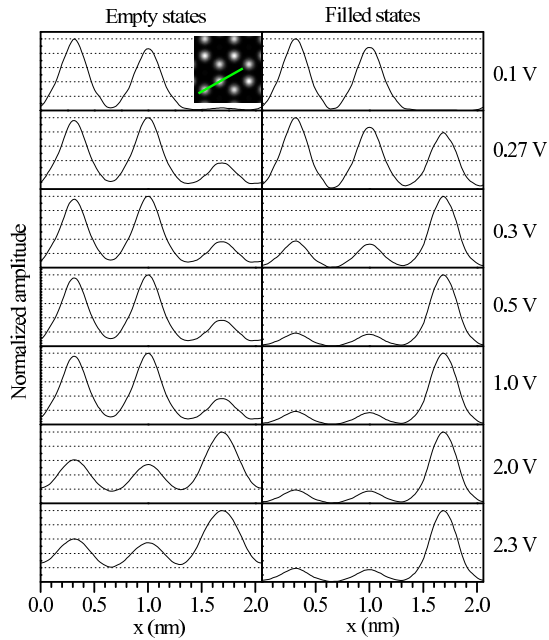}
\caption{\label{Fig4}(color online). Profiles of the simulated STM images for the 1U2D configuration along the line shown in the inset.}
\end{figure}
\\
The only remaining puzzling apparent contradiction comes from the Sn-4d core level spectroscopy that shows an opposite ratio of the two components with respect to the expected one. A recent paper \cite{Tejeda} suggests an explanation for such a mismatch, proposing that the two down adatoms are not equivalent. In that case the experimental core level spectra could be successfully fitted using three components in the deconvolution of the Sn-4d peaks. Actually, though the calculated height difference of the two down adatoms is negligible ($<$ 0.01 \AA ), the inequivalence of the two down Sn adatoms is confirmed by both ab-initio calculations and STM measurements. Indeed, the profiles (reported in Figure \ref{Fig4}) of the 1U2D simulated images show that in some cases the two down adatoms have a different apparent height. In STM measurements the inequivalence becomes evident at high bias voltage in filled states images. 
However, our calculations show that such inequivalence is very small, as demonstrated by the theoretical energy difference between the 4d core level peaks calculated for the two down Sn adatoms ($\Delta $E=0.01 eV, much smaller than reported in Ref. \cite{Tejeda}, $\Delta $E=0.16 eV).
\\
In conclusion, by comparing first-principles quasiparticle calculations and experimental STM results at the $\alpha $-Sn/Ge(111) surface, we obtained a clear indication about the configuration of such a system. All the evidences, namely the fact that first-principles calculations predict a stable system for the 1U2D configuration only and the comparison between bias-dependent simulated and experimental STM images, unambiguously indicate that the stable configuration for the $\alpha $-Sn/Ge(111) surface is the 1U2D.
Furthermore, our data confirm the inequivalence of the two Sn adatoms suggested by Tejeda et al. \cite{Tejeda}, although with a much smaller difference.
\\
We would like to warmly thank M. Luce and A. Ippoliti for their expert technical support.
CPU time at CINECA has been granted by CNR-INFM.
This work has been supported by the EU through the Nanoquanta Network of Excellence (NMP4-CT-2004-500198).

% If in two-column mode, this environment will change to single-column
% format so that long equations can be displayed. Use
%\begin{widetext}
%\end{widetext}

% figures should be put into the text as floats.
% Use the graphics or graphicx packages (distributed with LaTeX2e)
% and the \includegraphics macro defined in those packages.
% See the LaTeX Graphics Companion by Michel Goosens, Sebastian Rahtz,
% entire page. There is no need to do explicit centering.

% \begin{figure}
% \end{figure}

% Surround figure environment with turnpage environment for landscape
% figure
% \begin{figure}
% \end{figure}

%\newpage

\begin{thebibliography}{99}
\bibitem{det_dft} Norm-conserving LDA pseudopotentials have been employed, but ultrasoft gradient-corrected pseudopotentials have 
also been tested, giving comparable results. A kinetic energy cutoff of 12 Ry, increased to 20 Ry for convergence checking, 
has been used. The 3$\times $3 surface Brillouin zone (SBZ) has been sampled by a 
uniform mesh of 15$\times $15 k-points.
\bibitem{det_gw}  In GW calculations, the same mesh of k-points has been used as for DFT. A plasmon pole model has been applied to 
parametrize the dielectric function, that has been evaluated using 600 bands and 1087 plane waves. 1087 plane waves have also been used 
for the correlation part of the self energy and 12993 for the exchange part.
\bibitem{num_lay} A convergence check has been performed with respect to the number of Ge layers: relaxation of an eight-layer slab resulted
in a somewhat smaller tin adatoms ripple (0.28 \AA), but the overall picture is not altered.
\end{thebibliography}
\end{document}